\documentclass[prb,superscriptaddress,showpacs,twocolumn,longbibliography]{revtex4-1}

\usepackage{hyperref}

\usepackage{color}
\usepackage[usenames,dvipsnames]{xcolor}
\usepackage{amsmath,amsthm,amssymb}
\usepackage{graphicx}
\usepackage{epsfig}
\usepackage{dcolumn}
\usepackage{bm}
\usepackage{mathrsfs}
\usepackage{multirow}
\usepackage[all]{xy}
\usepackage{pbox}
\usepackage{verbatim}

\def\(({\left(}
\def\)){\right)}
\def\[[{\left[}
\def\]]{\right]}

\newcommand{\be}{\begin{equation}}
\newcommand{\ee}{\end{equation}}
\newcommand{\ben}{\begin{eqnarray}}
\newcommand{\een}{\end{eqnarray}}
\newcommand{\beq}{\begin{equation}}
\newcommand{\eeq}{\end{equation}}

\begin{document}

\title{Anderson and many-body localization in the presence \\of spatially correlated classical noise}

\author{Stefano Marcantoni}
\affiliation{School of Physics and Astronomy, University of Nottingham, Nottingham, NG7 2RD, UK}
\affiliation{Centre for the Mathematics and Theoretical Physics of Quantum Non-Equilibrium Systems, University of Nottingham, Nottingham, NG7 2RD, UK}

\author{Federico Carollo}
\affiliation{Institut f\"ur Theoretische Physik, Universit\"at T\"ubingen, Auf der Morgenstelle 14, 72076 T\"ubingen, Germany}

\author{Filippo M. Gambetta}
\thanks{Present address: Phasecraft Ltd., Bristol, UK.}
\affiliation{School of Physics and Astronomy, University of Nottingham, Nottingham, NG7 2RD, UK}
\affiliation{Centre for the Mathematics and Theoretical Physics of Quantum Non-Equilibrium Systems, University of Nottingham, Nottingham, NG7 2RD, UK}

\author{Igor Lesanovsky}
\affiliation{School of Physics and Astronomy, University of Nottingham, Nottingham, NG7 2RD, UK}
\affiliation{Centre for the Mathematics and Theoretical Physics of Quantum Non-Equilibrium Systems, University of Nottingham, Nottingham, NG7 2RD, UK}
\affiliation{Institut f\"ur Theoretische Physik, Universit\"at T\"ubingen, Auf der Morgenstelle 14, 72076 T\"ubingen, Germany}

\author{Ulrich Schneider}
\affiliation{Cavendish Laboratory, University of Cambridge, J. J. Thomson Avenue, Cambridge CB3 0HE, United Kingdom}

\author{Juan P.  Garrahan}
\affiliation{School of Physics and Astronomy, University of Nottingham, Nottingham, NG7 2RD, UK}
\affiliation{Centre for the Mathematics and Theoretical Physics of Quantum Non-Equilibrium Systems, University of Nottingham, Nottingham, NG7 2RD, UK}

\date{\today}

\begin{abstract}
    We study the effect of spatially correlated classical noise on both Anderson and many-body localization of a disordered fermionic chain. By analyzing the evolution of the particle density imbalance following a quench from an initial charge density wave state, we find prominent signatures of localization also in the presence of the time-dependent noise, even though the system eventually relaxes to the infinite temperature state. In particular, for sufficiently strong static disorder we observe the onset of metastability, which becomes more prominent the stronger the spatial correlations of the noise. In this regime we find that the imbalance decays as a stretched-exponential --- a behavior characteristic of glassy systems. We identify a simple scaling behavior of the relevant relaxation times in terms of the static disorder and of the noise correlation length. We discuss how our results could be exploited to extract information about the localization length in experimental setups.
\end{abstract}

\maketitle

\section{ Introduction} Understanding the dynamical behavior of closed many-body quantum systems is a subject of intense research. In particular the problem of thermalization \cite{Srednicki_1994,Rigol_2008, Yukalov_2011, DAlessio_2016, Gogolin_2016}, and the lack of it \cite{Basko_2006,Znidaric_2008,Bardarson_2012,Hauke_2015,Vosk_2015,Altman_2015,Nandkishore_2015,Luitz2015, Schreiber_2015,Bordia_2016,Smith2016,Choi2016, Imbrie_2017,DeRoeck_2017,Ho_2019,Abanin_2019,Suntajs_2020,Sels_2021,Serbyn_2021}, has given rise to a whole host of experimental and theoretical investigations over the past years. Among the mechanisms that prevent thermalization, a lot of attention has been devoted to many-body localization (MBL), that is the generalization of Anderson localization (AL) \cite{Anderson_1958} to disordered systems with interactions. Both localization phenomena have also been studied in the presence of periodically modulated noise  \cite{Delande_2017,Sacha_book} and local dissipation \cite{Levi_2016,Fischer_2016,Medvedyeva_2016,Luschen_2017,Lenarcic_2020,Ren_2020,Khosla_2022}.

\begin{figure}[t]
\centering
 \includegraphics[width=0.45\textwidth]{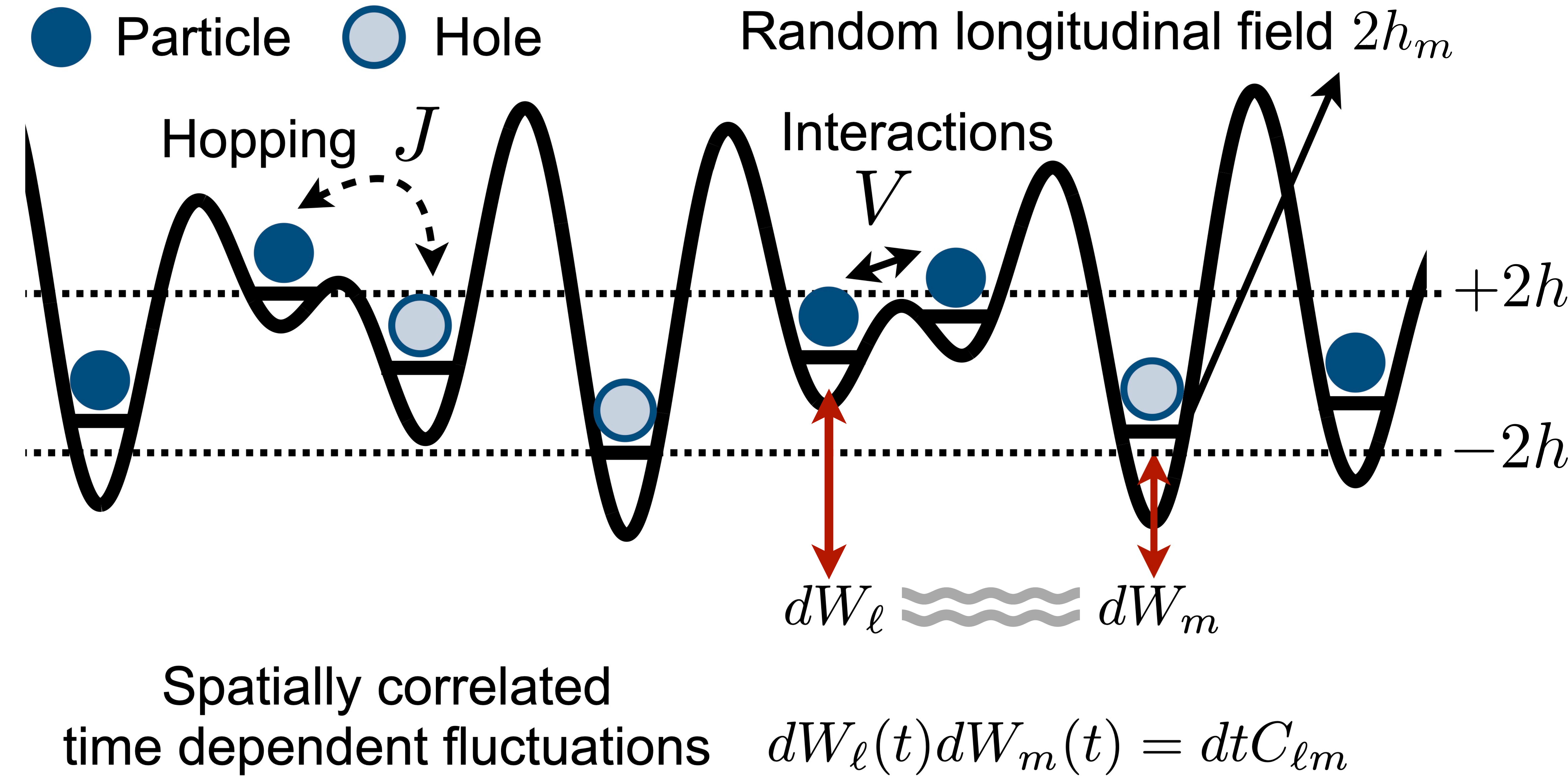}
\caption{{\bf Model.} Sketch of the model of fermions on a lattice with a random onsite potential uniformly distributed between $-2h$ and $2h$, hopping amplitude $J$, interaction  between neighboring sites $V$ and spatially correlated noise with correlation matrix $C_{\ell m}$. }
\label{Fig1}
\end{figure}

Motivated by the potential for experimental realisations, in this work we investigate the effects of \emph{spatially correlated} classical noise on AL and MBL. 
For our study we consider a fermionic chain with static disorder subjected to dynamical noise [cf.~Fig.~\ref{Fig1}]. The quantum state of this system, averaged over different realizations of the noise, evolves according to a Markovian master equation of the Gorini-Kossakovski-Sudarshan-Lindblad form \cite{GKS_1976,Lindblad_1976}.  We study the temporal evolution of the occupation imbalance between odd and even sites, after initializing the system in a charge density wave (or N\'eel) state.  This quantity has already been measured experimentally in cold atoms settings \cite{Schreiber_2015,Bordia_2016} and it is intuitively related to localization: at long times in a delocalized phase the imbalance vanishes, while it remains finite in a localized phase.  In our dissipative setting,  asymptotically in time,  the system relaxes to the infinite temperature state and, therefore, to zero imbalance. However, for both AL and MBL we observe metastable localization which manifests in a slow decay of the imbalance. 
Similarly to previous works studying MBL with local dephasing or different kinds of dissipative effects \cite{Levi_2016,Fischer_2016,Medvedyeva_2016,Luschen_2017,Lenarcic_2020,Ren_2020,Khosla_2022},
we find that this metastable localization relaxes in a stretched exponential manner, with a timescale that increases with increasing correlation length in the noise. Moreover, we show evidence of a scaling behavior in the relaxation time that could be used to extract experimentally a relative  localization length of two systems with different degrees of disorder,  at least in the non-interacting case. 

\section{ The model} We study here the time evolution of a one-dimensional fermionic chain subject to dephasing stochastic noise, as sketched in Fig.~\ref{Fig1}. The coherent part of the dynamics of the system is governed by the Hamiltonian
\begin{align}
H_{0}=\sum_{k=1}^{L-1}\!J\!\left(c_k^\dagger c_{k+1}+\text{h.c.}\right)+V\!\sum_{k=1}^{L-1} n_k n_{k+1}\!+2\sum_{k=1}^Lh_kn_k,
\label{MBL-Ham}
\end{align}
where $L$ is the length of chain, $c_k$ and $c_k^\dagger$ are fermionic annihilation and creation operators, respectively, and $n_k=c^\dagger_k c_k$ is the number operator at site $k$. The term proportional to the rate $J$ describes nearest-neighbor particle hopping. The remaining two terms, which are diagonal in the number-operator basis, instead describe interactions between fermions (the term proportional to $V$) and a random longitudinal field, with $h_k\in[-h,h]$, where the $h_k$ are uncorrelated and uniformly distributed quenched random variables. This system undergoes a MBL phase transition for large enough values of the disorder strength $h$ \cite{Vu_2022}. A similar phenomenon takes place also in the absence of interactions, $V=0$, where any amount of disorder $h$ is sufficient to produce AL. Features of these models and of the related phase transition have been widely investigated \cite{Znidaric_2008,Vosk_2015,Imbrie_2017,Vu_2022}, and some results can also be found in the presence of local dissipation \cite{Levi_2016,
Fischer_2016,Medvedyeva_2016,Luschen_2017,Ren_2020, Khosla_2022}. 

Here, instead, we consider the experimentally relevant case where the unitary evolution with the Hamiltonian $H_0$ is affected by classical dynamical noise in the local longitudinal fields. This noise can describe experimental imperfections or it can even be implemented experimentally on-demand using e.g.\ focused off-resonant light fields in quantum gas microscopes or ion traps, or local biases in superconducting circuits. To model such a dynamical noise, we make use of classical Wiener processes $dW_k(t)$ describing Markovian zero-average Gaussian fluctuations of the random potentials. While Markovianity states that the noise is not correlated in time, we can still account for equal-time spatial correlations by introducing a non-diagonal covariance matrix in the Ito rules \cite{Oksendal_book} describing the noise. Concretely, we consider the stochastic Hamiltonian dynamics with infinitesimal unitary evolution operator defined as $U_{t+dt,t}=\mathrm{e}^{-idH_t}$ with
\begin{equation}
dH_t=H_0\, dt +\sum_{k=1}^L n_k\, dW_k(t)\, .
\label{stochasticHamiltonian}
\end{equation}
We can model dynamical spatial correlations of the Wiener processes $dW_k(t)$ by the following Ito relations 
\begin{equation}
dW_k(t)dW_j(t)=C_{kj}\, dt,
\label{Ito-Rules}
\end{equation}
where $C\ge0$ is the covariance matrix of the classical noises. To enforce the positivity of this matrix and to have a direct control of the spatial correlation length of the noise we assume
$$
C_{kj}=\gamma \exp\left(-\frac{|k-j|}{\xi}\right)\, ,
$$
where $\xi$ is what we call correlation length and $\gamma$ is a coupling constant. Limiting cases are $\xi\to0$ for which the noise is spatially uncorrelated, $C_{kj}=\gamma \delta_{kj}$, and $\xi\to \infty$ for which the noise is infinitely correlated, $C_{kj}=\gamma$ for any pair $(k,j)$.\\

The resulting dynamics is a unitary stochastic evolution for the random vector $|\psi_t\rangle$, described by the stochastic differential equation \cite{Belavkin_1990, Barchielli_book} 
\begin{align}\label{sde}
 d |\psi_t\rangle =&-\Big(iH_0 +\frac{1}{2}\sum_{k,j=1}^L C_{kj} n_k n_j \Big)dt |\psi_t\rangle \nonumber\\
 &-i\sum_{k=1}^L n_k dW_k(t) |\psi_t\rangle.
\end{align}
A solution of the equation above represents a single dynamical realization (or trajectory) of the quantum time evolution for a fixed random profile $h_k$ of the longitudinal potential. However, typical order parameters for investigating localization phenomena can also be obtained from the pure state projector $|\psi_t\rangle \langle \psi_t|$ averaged over all realizations $\rho_t=\mathbb{E}[|\psi_t\rangle \langle \psi_t|]$, where $\mathbb{E}[\cdot]$ means expectation over the Wiener processes. 

One can obtain the evolution of $\rho_t$ using the Ito product rule $d |\psi_t\rangle\langle \psi_t | =  (d|\psi_t\rangle)\langle \psi_t | +  |\psi_t\rangle (d\langle \psi_t |) +  (d|\psi_t\rangle)(d\langle \psi_t |)$. Exploiting the evolution equation above, Eq.~\eqref{sde}, and the Ito relations Eq.~\eqref{Ito-Rules}, one arrives at
\begin{align}
&\dot{\rho}_t=-i[H_0,\rho_t]+ \mathcal{D}[\rho_t], \nonumber \\
& \mathcal{D}[\rho_t] =\sum_{k,h=1}^L C_{kj}\left(n_k\rho_t n_j-\frac{1}{2}\left\{n_j n_k,\rho_t\right\}\right)\, .
\label{Lindblad}
\end{align}
The resulting equation is in (non-diagonal) Lindblad form and it implements the time evolution of the average quantum state. The following analysis is based on the numerical study of this master equation.  Note that for $V=0$ the dynamics above maps the two-point functions $c_k^\dagger c_j$ onto two-point functions so that their evolution can be simulated efficiently. Such information is enough to investigate the density imbalance between odd and even sites (see Sec.~\ref{sec:noise_vs_localization}). \\

\begin{figure*}[t]
 \includegraphics[width=\textwidth]{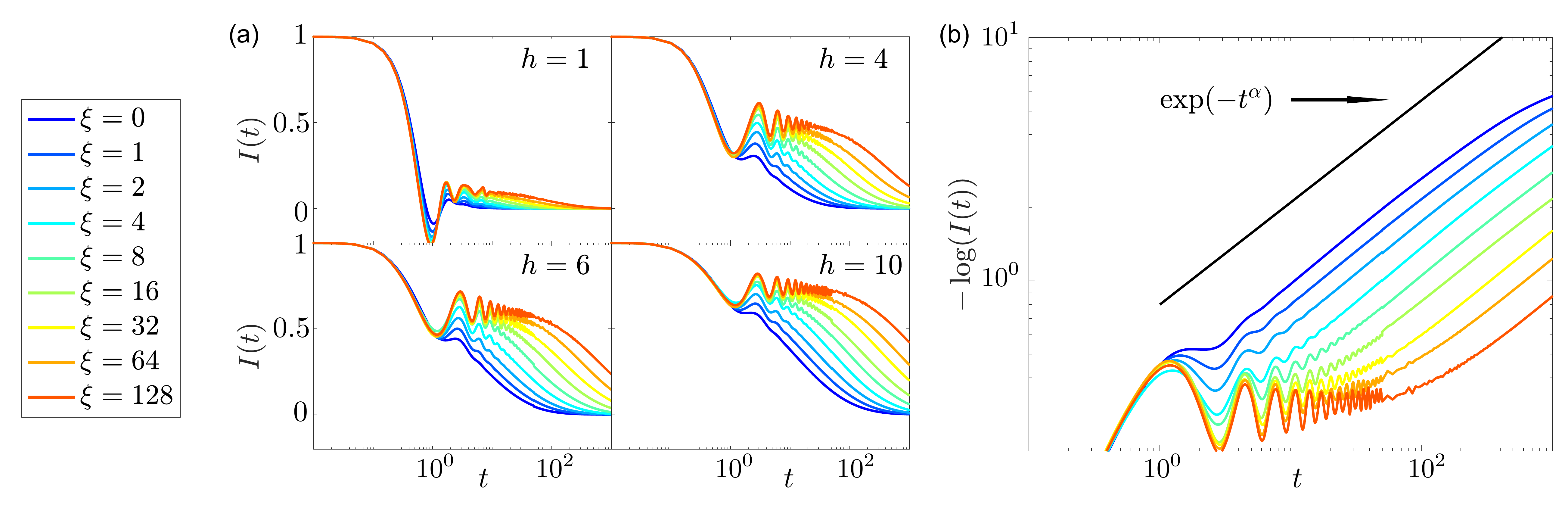}
\caption{{\bf Metastability of the imbalance.} (a) Imbalance versus time (note log scale on time axis), for a fixed dephasing rate ($\gamma=1$), and different values of $h$ and $\xi$. The metastable plateau is due to localization effects and it is higher (i.e., more memory about the initial condition is retained) for larger disorder strength.  The correlation length $\xi$ controls the duration of the plateau, with a longer plateau for larger $\xi$. The Hamiltonian case (stable plateau) is recovered for $\xi \to \infty$. (b)  The quantity $-\log(I)(t)$ is plotted for $h=10$ in log-log scale to highlight the stretched exponential decay of the imbalance. For the different values of $\xi$, the exponent $\alpha$ is approximately the same, $\alpha\approx 0.42$. (Results for $L=80$, averaged over $80$ disorder samples.)
}
\label{Fig2}
\end{figure*}

\section{Classical  noise versus localization}
\label{sec:noise_vs_localization}
We start by investigating the effects of correlated classical noise on the phenomenon of Anderson localization.  In particular, we consider the fermionic chain described by $H_0$, Eq.~\eqref{MBL-Ham} in the case of vanishing interactions between fermions, $V=0$.  In this regime, the system is expected to be localized for any strength of the disorder parameter $h$.  Even though Anderson localization is a single-particle effect, it shares many features with MBL. Therefore,  it is reasonable to expect that the impact of classical dynamical noise may be similar in many respects for AL and MBL.  

A simple quantifier of localization is provided by the evolution of the imbalance, $I(t)$, defined from observables as follows
\begin{equation}
I(t)=\frac{\langle N_{\rm odd}-N_{\rm even}\rangle }{\langle N_{\rm odd}+N_{\rm even}\rangle }\, ,
\label{imbalance}
\end{equation}
where $N_{\rm odd}$ ($N_{\rm even}$) is the number operator on odd (even) sites of the lattice. In order to avoid a priori asymmetries between even and odd sites,  we always consider $L$ even in the following.
Starting from a staggered initial state $\rho_0=|\psi_0\rangle \langle \psi_0|$ (N\'eel state), with $|\psi_0\rangle=\prod_{k=1}^{L/2}c^\dagger_{2k-1}|0\rangle$ and $|0\rangle$ the Fermi vacuum, one has $I(0)=1$. At later times, in the presence of localization, we find that the imbalance after a transient converges to a value different from zero. This indicates that some memory of the initial state is preserved by the existence of a characteristic localization length. In de-localized regimes instead the imbalance goes to zero, since particles initially trapped in odd sites travel freely in the system and, after some time, there is no possibility of recovering any information about the initial state.

 We compute the imbalance using numerical diagonalization of the dynamics, for a range of model parameters. In the non-interacting case, even if the Lindblad equation \eqref{Lindblad} is not quadratic in terms of fermionic creation/annihilation operators, one can show that the evolution of two-point operators $c_k^\dagger c_j$ is closed on two-point operators. This allows to perform the diagonalization only in the subspace of two-point operators and to access large system sizes $L$. All the following plots related to the Anderson case ($V=0$) are for $L=80$.  The hopping parameter $J$ is fixed to $1$ in the rest of the manuscript, or equivalently, all the other parameters in the generator are expressed in units of $J$. 

In Fig.~\ref{Fig2}, panel (a), we plot the imbalance as computed in the average state $\rho_t$ evolving through the Lindblad dynamics in Eq.~\eqref{Lindblad}, further averaged over $80$ different realizations of the random longitudinal fields $\{h_k\}_0^L$.  Each of the  pictures represents the imbalance for different values of the correlation length $\xi$ and a fixed strength of the disorder $h$, while $\gamma=1$ in all the figures.  For large enough disorder, a metastable plateau emerges (note the log scale in time), reminiscent of the localization effect in the corresponding Hamiltonian model.  Higher values of the disorder correspond to a higher plateau.  The length of the plateau is instead associated with the correlation length of the noise $\xi$.  In particular, this highlights how, for increasing $\xi$, the non-ergodic behavior of AL is preserved for longer and longer times. This seems to be due to the fact that, for correlation lengths that are larger or comparable to the localization length, dephasing effects require more time to become effective since the dynamics within the correlation length of the noise is almost equivalent to the deterministic localized dynamics.  

\begin{figure*}[t]
\centering
\includegraphics[width=\textwidth]{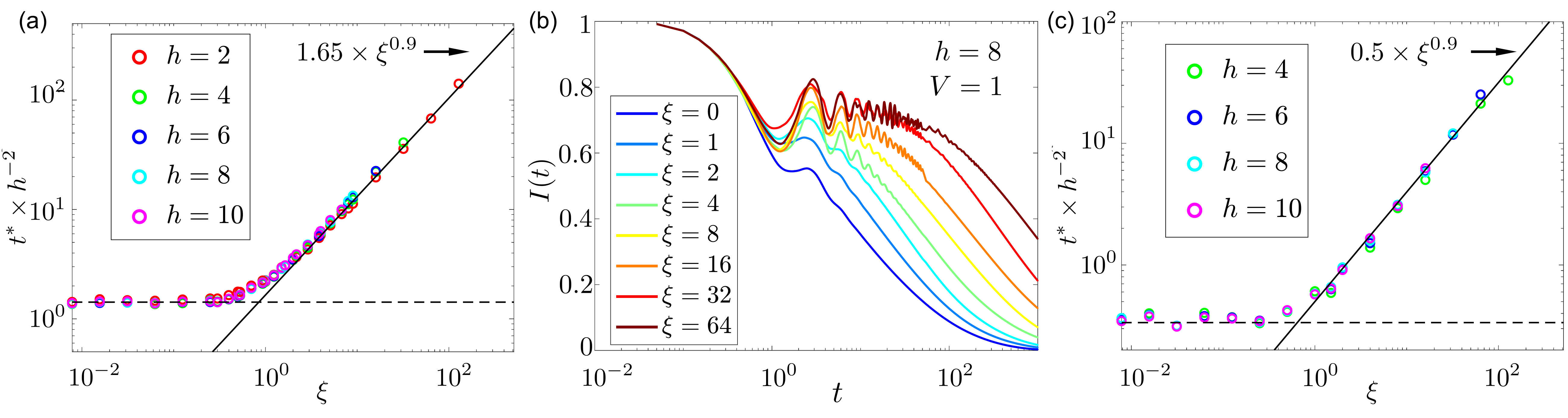}
\caption{{\bf (a) Scaling of the relaxation time (no interaction)}. Time $t^*$  to reach imbalance $I=0.05$,  plotted as a function of $\xi$ for different values of $h$. Rescaling $t^*$ by a factor $h^2$ results in collapse onto a single curve. The behavior is flat for values of $\xi$ too small to see the lattice spacing, while it is power-law ($\sim \xi^{0.9}$) for large $\xi$. The collapse of curves shows that one has $t^* \sim h^2 \xi^{0.9}$ at large $\xi$. {\bf (b) Metastability with interactions}.  Imbalance versus time, for fixed dephasing rate ($\gamma=1$), fixed disorder strength ($h=8$), and interaction ($V=1$), for different values of $\xi$.  {\bf (c) Relaxation time with interactions}. The relaxation time $t^*$ (with threshold imbalance $I=0.25$) is plotted against $\xi$ for different values of $h$ and $V=1$. All curves, after rescaling by $h^2$, collapse onto a single one, thus showing that at large $\xi$ one has $t^* \sim h^2 \xi^{0.9}$. }
\label{Fig3}
\end{figure*}

In the non-interacting case we are considering, this effect can also be understood by looking at the action of the dissipative part of the generator on two-point operators. Coherences between site $j$ and site $k$ are in this case embodied by the operator $c^\dagger_k c_j$ (and its Hermitian conjugate);  the action of the dissipative part of the Lindblad generator $\mathcal{D}$ on this is given by a damping term 
\begin{equation}
\mathcal{D}[c^\dagger_k c_j]\propto -{\gamma_{kj}}c^\dagger_k c_j,  \quad
\gamma_{kj}=C_{kk}+C_{jj}-2C_{kj}\, .
\end{equation} 
Therefore, when $\xi\gg|k-j|$ one has that $\gamma_{kj}\ll 1$, meaning that the destructive interference effects leading to AL are preserved for longer and longer times for spatially correlated noise. 
In the $\xi \to \infty$ limit, one recovers the Hamiltonian behavior.

After the metastable plateau, the imbalance asymptotically decays to zero, as the  dynamics \eqref{Lindblad} has a unique asymptotic state in each particle number sector, namely the infinite temperature state. However, the effect of AL can be seen also at this later stage, because the decay is slower than exponential for high enough disorder. In particular, as in Ref.~\cite{Levi_2016}, the imbalance follows a stretched exponential behavior 
$
I(t)\propto e^{-\mu t^\alpha}\, ,
$ 
with $0<\alpha<1$, $\mu>0$.
Panel (b) in Fig.~\ref{Fig2} shows the stretched exponential decay for the case $h=10$.  The parameter $\mu$ depends on the correlation length, while the exponent $\alpha$ in this case is approximately the same for all curves ($\alpha\approx 0.42$).  We note that the stretched exponential behavior of the imbalance is also observed at the level of single realizations of the random longitudinal field. This means that such non-exponential decay is not due to an average over different instances of such disorder, rather it may be the result of a self-averaging mechanism emerging in single realizations of the field, for sufficiently large $L$. 

We can further analyze the slow decay of the imbalance by looking at the \emph{relaxation time} $t^*$ needed to reach a certain threshold value as a function of $h$ and $\xi$.  For small enough $\xi$ the noise is completely uncorrelated until a value comparable to the lattice structure is reached.  Then, the relaxation time starts to increase and, for large enough $\xi$, it goes as a power law.  

The curves look very similar for different values of $h$.  Indeed,  as shown in Fig.~\ref{Fig3}(a) all the curves can be superimposed by means of a suitable rescaling, i.e.\ by dividing $t^*$ by $h^2$.
In the absence of dissipation, the Hamiltonian of the non-interacting system presents exponentially localized eigenfunctions, each one with a characteristic localization length. For a generic wave function or density matrix,  one may try to define an energy-averaged localization length.  However,  even without specifying a precise value,  for the range of disorder we study, the localization lengths of the unitary system are known to scale as $\propto h^{-2}$ (this can be shown with analytical perturbative calculations for small disorder \cite{Casati_1992} but appears to be valid numerically even for $h \sim 10$ \cite{Varga_1994}). 
Therefore, the observed scaling can be used to extract the value of $h^2$ by measuring the relaxation time for different values of the correlation length $\xi$. This can also provide relative information about the localization length.

A similar analysis can be done in the presence of interactions $V\neq0$.  The system sizes we can reach in this case are much smaller, because the space of two point correlation functions is not closed under the dynamics, so that one has to diagonalize the full Lindbladian. However, already for $L=6$, averaging over 100 realizations of the static disorder, we can see qualitatively that a similar situation emerges as in the AL case.  Fig.~\ref{Fig3}(b) shows the behavior in time of the imbalance for $V=1$, $h=8$ for different values of the correlation length $\xi$. Just like in the AL case, our results show evidence that
also in the MBL case metastable localization persists for longer times the larger the spatial correlations in the noise.

One can study also in the MBL case the relaxation time towards stationarity, see Fig.~\ref{Fig3}(c).  As in the AL case, the behavior for large $\xi$ is power law with almost the same power for different values of $h$.  The scaling of the curves with $h^2$ works also in the MBL case.

\section{Experimental setup}In addition to help in understanding the effects of experimental noise, the observed scaling also implies that adding correlated dynamical noise and measuring $t^*$ as a function of $\xi$ can be used to infer information about the effective localization length of a disordered system. Experimentally, this requires one to add well-characterized dynamical noise with varying correlation lengths. This can be straightforwardly implemented in most modern platforms. In the case of quantum gas microscopes \cite{Bakr_2009,Sherson2010d,Cheuk_2015, Gross_2021}, for instance,  one could use tightly focused off-resonant laser beams and the correlation length of the noise could be controlled by varying the focus size. One possible implementation would be based on injecting acousto-optical deflectors with multiple radio frequencies, similar to techniques developed in the context of optical tweezer arrays \cite{Bernien_2017,Kim_2018,Lorenz_2021,Scholl_2021,Kaufman_2021}.

\section{ Summary }
We have studied the fate of disorder-induced localization in the presence of spatially correlated classical noise. In the non-interacting case, where large system sizes can be studied numerically, we found that the metastable localization regime becomes longer in time with both  disorder strength and correlation length of the noise. In the interacting case a similar phenomenology is visible, even though only small sizes can be studied numerically. In both cases we found a simple scaling form for the relaxation time to stationarity. 
The physics studied could be readily investigated experimentally with the proposed setup.

\noindent {\bf \em Acknowledgments -- } This work was supported by the EPSRC through the Grant No. EP/R04421X/1. S.M.  acknowledges the use of Augusta, the HPC facility of the University of Nottingham, for the numerical calculations. S.M. is grateful to Loredana M. Vasiloiu for useful comments on a preliminary version of the draft. I.L. and J.P.G. acknowledge funding from the Baden-W\"urttemberg Stiftung through project No.~BWST\_ISF2019-23. I.L. also acknowledges funding from the German Research Foundation (DFG) through the Research Unit FOR 5413/1, Grant No. 465199066. This work was supported by the University of Nottingham and the University of Tübingen's funding as part of the Excellence Strategy of the German Federal and State Governments, in close collaboration with the University of Nottingham. U.S. acknowledges funding from the European Commission (ERC Starting Grant QUASICRYSTAL) and EPSRC (EP/R044627/1 and EP/P009565/1).

\end{document}